\documentclass[sigconf]{acmart}

\AtBeginDocument{%
  \providecommand\BibTeX{{%
    Bib\TeX}}}

\usepackage{algorithmic}
\usepackage{graphicx}
\usepackage{textcomp}
\usepackage{xcolor}
\usepackage{wrapfig}
\usepackage{textcomp}
\usepackage{xcolor}
\usepackage{soul}
\usepackage{braket}
\usepackage[ruled,vlined, linesnumbered]{algorithm2e}

\def\BibTeX{{\rm B\kern-.05em{\sc i\kern-.025em b}\kern-.08em
    T\kern-.1667em\lower.7ex\hbox{E}\kern-.125emX}}

\begin{document}

\title{Quantum Data Breach: Reusing Training Dataset by Untrusted Quantum Clouds}











\author{Suryansh Upadhyay}
\affiliation{%
  \institution{The Pennsylvania State University}
  \streetaddress{University Park}
  \city{PA}
  \country{USA}}
\email{sju5079@psu.edu}

\author{Swaroop Ghosh}
\affiliation{%
  \institution{The Pennsylvania State University}
  \streetaddress{University Park}
  \city{PA}
  \country{USA}}
\email{szg212@psu.edu}







\begin{abstract}


Quantum computing (QC) has the potential to revolutionize fields like machine learning, security, and healthcare. Quantum machine learning (QML) has emerged as a promising area, enhancing learning algorithms using quantum computers. However, QML models are lucrative targets due to their high training costs and extensive training times.
The scarcity of quantum resources and long wait times further exacerbate the challenge. Additionally, 
QML providers may rely on a third-party quantum cloud for hosting the model, exposing the models and training data. As QML-as-a-Service (QMLaaS) becomes more prevalent, reliance on third party quantum clouds can pose a significant threat. This paper shows that adversaries in quantum clouds can use white-box access of the QML model during training to extract the state preparation circuit (containing training data) along with the labels. The extracted training data can be reused for training a clone model or sold for profit. We propose a suite of techniques to prune and fix the incorrect labels. Results show that $\approx$90\% labels can be extracted correctly. The same model trained on the adversarially extracted data achieves approximately $\approx$90\% accuracy, closely matching the accuracy achieved when trained on the original data. To mitigate this threat, we propose masking labels/classes and modifying the cost function for label obfuscation, reducing adversarial label prediction accuracy by $\approx$70\%.

\end{abstract}





\maketitle

\section{Introduction}

Quantum computing (QC) has garnered significant interest due to its potential to revolutionize problem-solving across numerous fields. Leveraging principles such as superposition, entanglement, and interference, quantum computers promise exponential speedup for specific tasks compared to classical computers. With potential applications in machine learning \cite{cong2019quantum,biamonte2017quantum}, security \cite{ghosh2023primer,upadhyay2022robust}, drug discovery \cite{cao2018potential}, optimization \cite{farhi2014quantum,tilly2022variational}, finance \cite{orus2019quantum}, material science \cite{bauer2020quantum}, and healthcare \cite{gupta2023quantum}, quantum computing is becoming increasingly important in both academia and industry. Current quantum devices, categorized as Noisy Intermediate-Scale Quantum (NISQ), are limited by qubit connectivity and gate fidelity, which can result in incorrect output sampling. Due to these limitations, hybrid algorithms like the Quantum Approximate Optimization Algorithm (QAOA) and Variational Quantum Eigensolver (VQE) are being explored, using classical computers to iteratively adjust quantum circuit parameters. In this emergent field, quantum machine learning (QML) has gained considerable attention, aiming to improve learning algorithms by leveraging quantum capabilities. Various QML models have been explored, including quantum support vector machines (QSVMs) \cite{rebentrost2014quantum}, quantum convolutional neural networks (QCNNs) \cite{cong2019quantum}, and quantum generative adversarial networks (QGANs) \cite{lloyd2018quantum}. Among these, quantum neural networks (QNNs) \cite{abbas2021power} are notable for replicating the structure and function of classical neural networks within a quantum framework. QNNs rely on parameterized quantum circuits (PQCs) with trainable single-qubit and two-qubit gates, whose parameters are updated using classical optimizers. Optimizing PQCs in QNNs is challenging due to time, complexity, and quantum resource requirements. The no-cloning theorem prevents duplicating quantum states, making traditional backpropagation impractical. Instead, methods like the parameter shift rule are used, requiring multiple circuit executions to estimate gradients, which significantly increases resource demand as the number of parameters grows.

\textbf{Motivation:} Security and reliability are crucial in quantum computing, particularly with cloud-based access. Platforms from companies like like IBM \cite{IBMQuantum}, Google \cite{GoogleQuantumComputer}, and AWS Braket \cite{AmazonBraket} offer convenience and scalability but struggle with job submission latency, queue backlogs, and high costs. As the quantum computing ecosystem evolves, third-party service providers are expected to emerge, offering potentially higher performance at cheaper prices, enticing users to utilize these services. Examples include Orquestra\cite{Orquestra} and tKet \cite{pytket}, which support multiple hardware vendors. Baidu's "Liang Xi" \cite{baidu} solution provides access to various quantum chips via mobile app, PC, and cloud, offering flexible quantum services through private deployment, cloud services, and hardware access. While trusted hardware is preferred for applications with significant economic or social impact, hybrid quantum-classical algorithms face substantial costs and delays due to the high number of iterations required. Governments and large entities may have dedicated resources, but these are expensive and geographically limited. This dependence on third-party compilers, hardware suites, and service providers raises reliability and security concerns. As QML expands, we will see a rise in hosting uniquely architected QML models or those trained on novel quantum datasets on the cloud, leading to Quantum Machine Learning-as-a-Service (QMLaaS) \cite{MLaaS}. Deploying QML algorithms on untrusted, less-trusted or unreliable cloud-based quantum computing services exposes them to various adversarial attacks. These security risks arise due to several factors:

\textbf{High training cost:} Quantum computers are expensive, e.g., \$1.6 per second for IBM’s superconducting qubits and \$0.01 per shot for IonQ’s Trapped Ion (TI) qubits, which is at least $10^5 \times$ costlier than classical resources priced at $\sim\$ 2.1e-06$ per second. QML models require hundreds of training epochs, each with thousands of quantum circuit executions, making the trained and even partially trained QML model very expensive. Compared to current state-of-the-art ML models like Gemini that cost millions to billions of dollars for training, QML models at scale may cost many orders of magnitude higher, making them extremely valuable. 

\textbf{High training time:} Current state-of-the-art ML models e.g., ChatGPT3 took $\sim 1$ month for training using thousands of dedicated GPUs. Quantum resources are scarce, and their demand is extensive. Both hardware and simulators hosted in the cloud incur long wait queues, even with dedicated access like Quantum Hub memberships. Training a large QML model might take significant time (e.g., months to years).

\textbf{QMLaaS:} Since QML providers may not possess their own quantum hardware, they may rely on a third-party quantum cloud for hosting the model. This will lead to the rise of QMLaaS \cite{MLaaS} providing access to clients only through input-output queries via external APIs. The quantum cloud provider may have white-box access to the expensive model and training data. 

\textbf{Miscellaneous Intellectual Properties (IPs):} The untrained QML IPs include model architecture and training data embedded in state preparation circuit. The trained QML IPs include optimized parameters and input data embedded in the state preparation circuit during inference. 

Several studies have addressed protecting quantum circuits from
untrusted clouds \cite{upadhyay2022robust, upadhyay2023trustworthy}. However, there is a notable gap in research on the exploitation and protection of training data encoded in Quantum Neural Networks (QNNs) from cloud-based adversaries during training or inference stages. \textit{This paper to our knowledge is the very first attempt which aims to understand and mitigate this security vulnerability}. We address an attack where adversaries exploit the QNN to recover encoded data (in form of state preparation circuit) along with the labels from QNN model being trained on quantum clouds. The ability to reverse-engineer these encoded data points can enable counterfeit training dataset which can used for variety of purposes including effective cloning of the QML model. To defend against the adversarial attack, we propose two methods to secure the training data: (1) introducing masking labels/classes and (2) configuring the model so that only a specific combination of selected qubits yields the correct labels. We propose using a modified cost function that combines the standard loss function with an additional loss term that promotes incorrect label prediction if the measurement is taken from all qubits or any combination other than the user-selected qubits.


\textbf{Contributions:} In this work, we: (a) propose an attack model where an adversary in a quantum cloud can extract the state preparation circuit and labels for profit, (b) demonstrate the attack's effectiveness with various label extraction methods, (c) introduce two defensive methods: masking labels/classes and a modified cost function, and (d) validate our methods through simulations.

\textbf{Paper organization:} Section II provides background information. Section III outlines the threat model and section IV presents the proposed defense. Section V covers results and discussion and Section VI concludes the paper.

\begin{figure*}
    \centering
    \includegraphics[width= 7in]{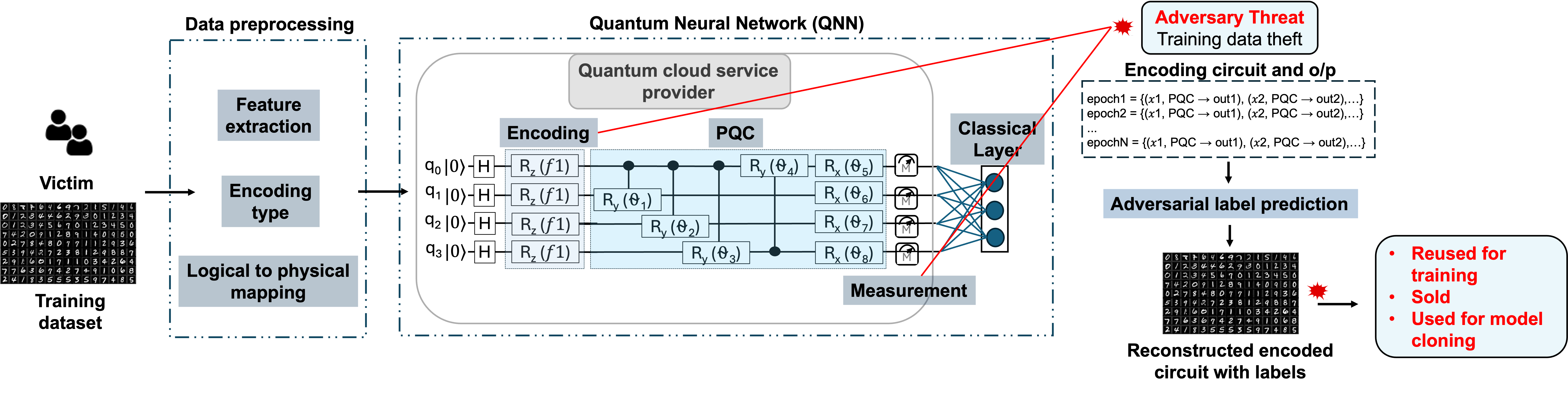}
    \caption{Proposed attack model where the adversary, posing as a reliable quantum cloud service provider, uses the white box access to the QNN submitted to a quantum cloud containing the state preparation circuit and the corresponding measurement results to extract the correct labels. 
    This dataset is then reused or sold.}
    \label{threat model}
\vspace{-4mm}
\end{figure*}

\section{Background}

\subsection{Quantum Computing}

Qubits are similar to classical bits in that they store data through internal states such as $\ket{0}$ and $\ket{1}$. However, due to their quantum nature, qubits can exist in a superposition of both $\ket{0}$ and $\ket{1}$. A qubit's state, denoted by $\varphi$ = a $\ket{0}$ + b $\ket{1}$, can be expressed as a combination of complex probability amplitudes, $a$ and $b$, corresponding to the states $\ket{0}$ and $\ket{1}$. When measured, it collapses to a single state, either $\ket{0}$ or $\ket{1}$, with probabilities of $|a|^2$ and $|b|^2$, respectively. 
Qubits are highly sensitive to noise and are error-prone. Key error types that affect quantum computing operations include: coherence errors\cite{iverson2020coherence}, operational errors\cite{magesan2012efficient}, measurement error\cite{busch2014colloquium} and crosstalk\cite{ash2020analysis}. 
\vspace{-3mm}

\subsection{Quantum Neural Network (QNN)}

A QNN is comprised of three main components (Fig. \ref{threat model}): (i) a circuit for encoding classical data into quantum states, (ii) a parameterized quantum circuit (PQC) with tunable parameters, and (iii) measurement operations. Due to qubit limitations, classical pre-processing techniques like Principal Component Analysis (PCA) are typically used to reduce the input data dimension. To encode reduced classical data into quantum states, various techniques such as basis encoding and amplitude encoding can be used \cite{abbas2021power}. For continuous variables, angle encoding is most common, where a classical feature is represented as a rotation of a qubit along a specific axis. Since qubit states repeat in $2\pi$ intervals, features are typically scaled between $0$ and $2\pi$ (or $-\pi$ to $\pi$) during pre-processing. The PQC consists of a sequence of quantum gates with adjustable parameters to solve a specific problem. 
The PQC includes entangling operations, which are multi-qubit operations (parameterized or not) to create correlated states, and parameterized single-qubit rotations to explore the solution space. Finally, measurement operations collapse the qubit states to either $0$ or $1$. The expectation value of Pauli-Z is used to determine the average state of the qubits. These measured values are typically fed into a classical neuron layer, with the number of neurons equal to the number of classes in the dataset. This layer performs the final classification task. A classical optimizer optimizes the parameters iteratively to achieve the desired input-output relationship.
\vspace{-3mm}
\subsection{Related works}
Secure computation techniques, such as homomorphic encryption and secure multiparty computation, protect machine learning processes by performing computations on encrypted data\cite{gilad2016cryptonets,peng2023rrnet} ensuring that outsourced computation and data remain secure in the cloud. However, these classical computing methods may not be applicable in quantum scenarios due to the unique challenges posed by quantum computations. For instance, quantum homomorphic encryption (QHE) theoretically allows computation on encrypted quantum data but is impractical due to its high computational overhead\cite{fisher2014quantum}. Research in quantum security has addressed model theft \cite{wang2023qumos,kundu2024stiq} and threats from untrusted hardware providers \cite{upadhyay2022robust,upadhyay2023trustworthy}, but unrelated to training data stealing attacks. QuMoS\cite{wang2023qumos} defends against model stealing by distributing model parts across isolated quantum cloud providers, limiting adversary access however it fails to protect training data. STIQ\cite{kundu2024stiq} uses an ensemble-based strategy to obfuscate QNN outputs on untrusted platforms, combining them locally to reveal accurate results. Unlike these methods which focus on model stealing and defense, our work demonstrates how adversaries can extract state preparation circuits and labels from QML models, effectively stealing training data. 
Strategies like dividing circuit execution between trusted and untrusted providers \cite{upadhyay2022robust,upadhyay2023trustworthy} are ineffective for QNNs because access by any single untrusted provider would expose the subsequent training data to the adversary.
\vspace{-2.5mm}
\section{Threat model}

\subsection{Basic Idea}

The assets in QNNs include proprietary algorithms, training data, and the resulting trained models. 
In the proposed attack model, the adversary takes the form of a less reliable or untrusted quantum service provider while posing as a reliable and trusted hardware provider. The adversary has access to the QNN submitted to a quantum cloud during training containing the state preparation circuit/encoding circuit and the corresponding measurement results (Fig. \ref{threat model}). The primary objective of the adversary is to extract the victim's state preparation circuit (which represents training data) and the corresponding labels. This data can then be sold for profit. Note, state preparation circuit is relatively easy to identify and extract from the QNN however, labels (that is used to compute the loss at the user end) for the corresponding data is not available to the quantum cloud. Adversaries can be driven by several motivations to steal data from QNNs, such as:

1) Access to proprietary training data gives adversaries a competitive edge, allowing them to develop or enhance models without the costs of data collection and processing.

2) The high cost and scarcity of quality training data make it a lucrative target. By stealing rare datasets, adversaries can cut costs and profit by selling the data to third parties.
\vspace{-2mm}
\subsection{Adversary Capabilities}

We assume that adversary has, \textbf{(a) access to the encoding circuit and results:} the adversary has access to the encoding circuit submitted by users and the results produced by the quantum computer. This is justified because the cloud provider, by design, has access to both the QNN and the measurement results, 
and \textbf{(b) epoch distinguishment:} adversary can distinguish between different epochs. Given access to the encoding circuits of the QNN, it is easy to identify repetitive patterns or input data points. 

To showcase the effectiveness of the attack, we also assume that adversary has knowledge about the classical pre-processing (e.g., PCA), the encoding technique (e.g., angle encoding) and logical to physical mapping of classical data to qubits that has been used to prepare the state preparation circuit. This is a strong assumption (as discussed in Section 5.3) and devoted effort is required from adversarial standpoint to extract this information from various features of the state preparation circuit. This study is outside the scope of the current research.  
\vspace{-2mm}
\subsection{Attack Process}

\textbf{(a) Data Collection Across Epochs:} The adversary collects state preparation circuits and corresponding predictions from all epochs. Each epoch involves feeding the PQC with encoded data and recording the output. For example, the adversary's visibility is illustrated in Fig. \ref{attack}.
The adversary aggregates data points ($x_1$,$x_2$....) and their corresponding predictions from all epochs into a comprehensive dataset. Each input point (encoded data in state preparation circuit) is paired with the expectation value or label (after being fed into a classical linear layer that produces the softmax probability vector). This aggregated dataset provides a complete view of the input-output relationships across all epochs, facilitating the extraction of meaningful labels (Fig.\ref{threat model}).

\textbf{(b) Label Determination Heuristics:} To identify the most likely correct label for each data point, the adversary can utilize several heuristics. In this work, we explore three methods:

\begin{figure}
    \centering
    \includegraphics[width= 3.5in]{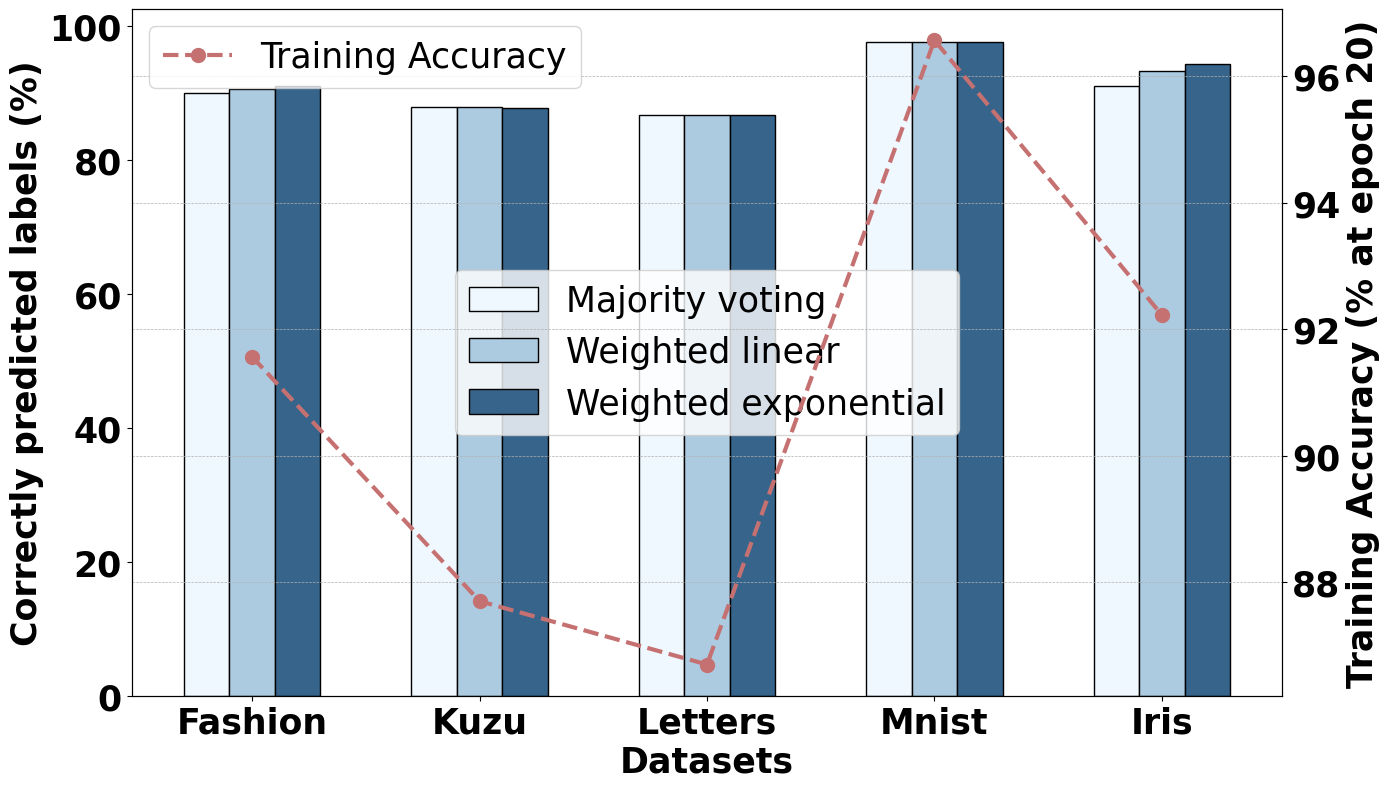}
    \caption{Performance comparison of label determination heuristics across five datasets using QNN models trained on a noiseless simulator for 20 epochs. }
    \label{correct_labels}
\vspace{-4mm}
\end{figure}

\textbf{1) Majority Voting}: If a data point receives multiple labels across epochs, the label with the highest frequency is selected. \textit{Example}: If a data point is labeled as A in 3 out of 5 epochs and B in 2, we select A as the final label.

\textbf{2) Weighted Majority Voting (Linear Weights)}: This method assigns linearly increasing weights to labels from successive epochs, prioritizing predictions from later epochs to enhance accuracy. \textit{Example}: Suppose a data point is labeled A in the 1st epoch, B in the 2nd, and B in the 3rd. With linear weights (1, 2, 3), the scores are A = 1 and B = 5, so the final label will be B.

\textbf{3) Weighted Majority Voting (Exponential Weights)}: This method uses exponentially increasing weights for labels from successive epochs, prioritizing predictions from later epochs. This approach intuitively follows the training curve of the QML models, where accuracy typically rises exponentially during initial epochs and then stabilizes. To reflect this, weights increase exponentially up to a rollover epoch, after which all weights remain constant. \textit{Example}: Suppose a data point is labeled as A in the 1$^{st}$ epoch, B in the 2$^{nd}$ epoch, and B in the 3$^{rd}$ epoch. With exponential weights (e.g., 1, 2, 4) up to the rollover epoch, the weighted scores are: A = 1, B = 2 + 4 = 6. After the rollover epoch, if the labels continue to be B, the weights assigned to further epochs remain fixed. Therefore, the final label is B. \textit{In this work we use the 90$^{th}$ percentile epoch as the rollover epoch.}

\textbf{(c) Creating a Refined Dataset:} The process of refining the dataset involves an iterative approach where we utilize k-fold cross-validation to identify and correct/prune mislabeled data points. The methodology is detailed as follows:

\textbf{K-Fold Cross-Validation:} We use this approach with different values of k to evaluate the performance of multiple classifiers \cite{anguita2012k}. For this work, \textit{we apply k-fold cross-validation with 4 values of k(5,7,10,15)}. For example, with $k=5$, the dataset is divided into subsets A, B, C, D, and E. In each iteration, one subset is used for validation while the remaining four are used for training, ensuring each data point is validated multiple times. Specifically, we train on subsets B, C, D, and E while testing on subset A, and repeat this process for all subsets. This approach ensures that each data point is validated multiple times, providing a robust mechanism for detecting mislabeled data points. We employ an ensemble of classifiers, including RandomForestClassifier, LogisticRegression, SVC, and MLPClassifier. Each classifier contributes to predicting the probability of each class for the data points.

\textbf{Mislabeled Point Identification:}  For each iteration, we aggregate the predicted probabilities for all k values for each classifier and determine the final predicted class for each data point. If the predicted class differs from the actual class, the data point is marked as potentially mislabeled only if it differs across all the classifiers.

\textbf{Updating and Pruning:} Data points identified as potentially mislabeled are subjected to further scrutiny. If the confidence (probability) of the predicted label exceeds a predefined threshold, the label is updated to the predicted label otherwise, the data point is pruned from the dataset. \textit{Example}: If a data point's predicted label has a confidence of 0.7 and the threshold is 0.6, the label is updated; otherwise, the point is pruned. \textit{In this work we use a threshold of 0.8.}

\textbf{Iteration and Refinement:} The updated dataset is then used in the next iteration of k-fold cross-validation, repeating the processes of identification, updating, and pruning. This iterative approach gradually refines the dataset by eliminating or correcting mislabeled points.

\begin{figure}
    \centering
    \includegraphics[width= 3.5in]{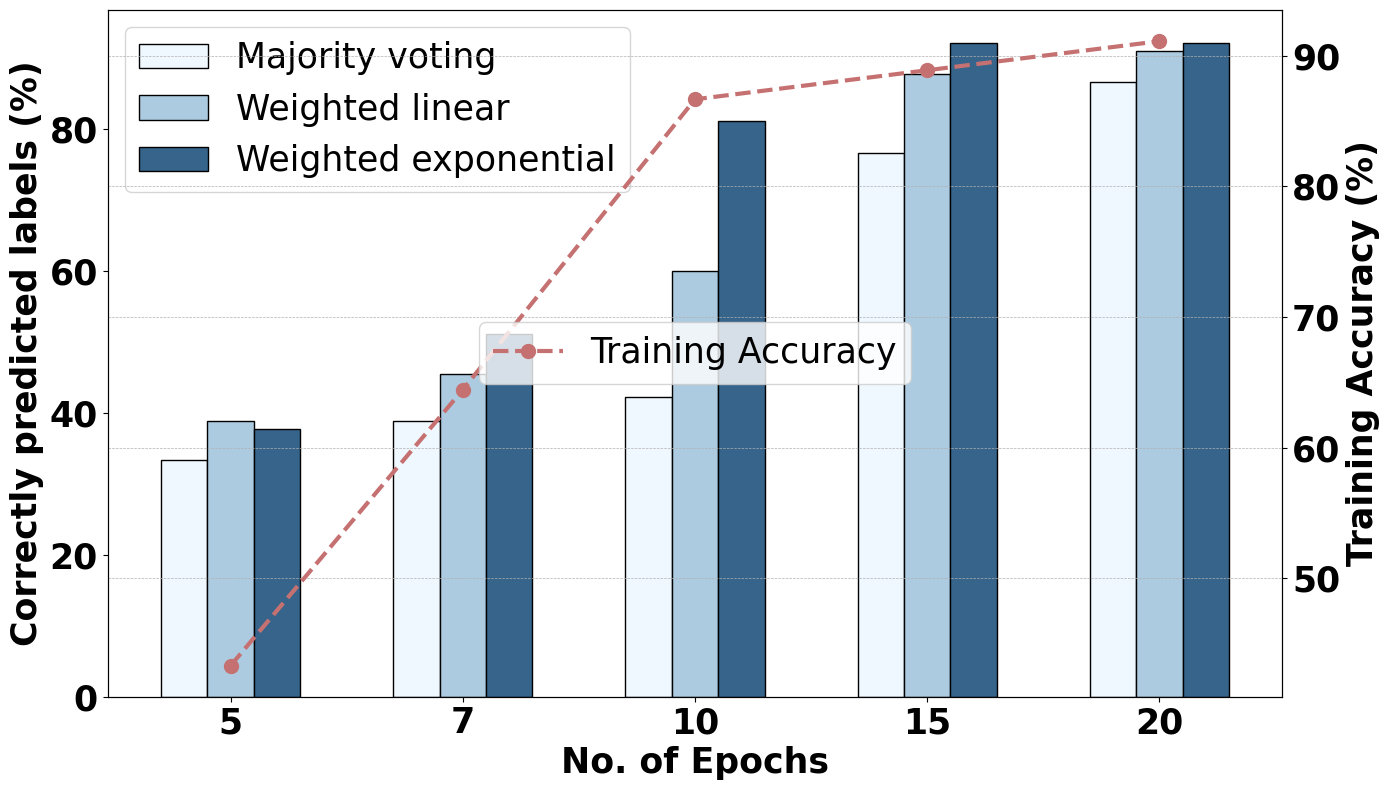}
    \caption{Comparison of label prediction accuracy across various heuristics for varying training accuracies using the Iris dataset on the backend device Fake27QPulseV1.}
    \label{iris_epochs}
\vspace{-4mm}
\end{figure}


\section{Proposed defense}
To defend against the proposed attack model, we employ two methods to secure the training data: 

\subsection{Introducing masking labels/classes}

In typical QNN setups, the user measures the number of qubits corresponding to the exact classes. For instance, in a 4-qubit QNN for a 3-class dataset, only 3 qubits are measured, with each qubit corresponding to a class. However, if all qubits are measured, the adversary in our attack model setting would be unable to determine the number of classes and might predict an incorrect label for a given data point. Leveraging this, we train the QNN with additional masking labels such that correct label prediction is achieved only when measurements from the specific qubits known only to the user are considered. This dual-layer masking ensures that the adversary cannot discern the actual number of classes or which specific qubit measurements to consider. For example, if a 4-qubit QNN is trained for a 3-class dataset with a masking label such that the correct label is only predictable when measuring qubits 1, 2, and 3, an adversary considering all 4 qubits/any other qubit combination will not be able to accurately predict the correct class for the data points. Consequently, the adversary, without knowledge of true number of classes and which specific qubits to consider, will not be able to accurately predict the labels.

\begin{figure}
    \centering
    \includegraphics[width= 3.5in]{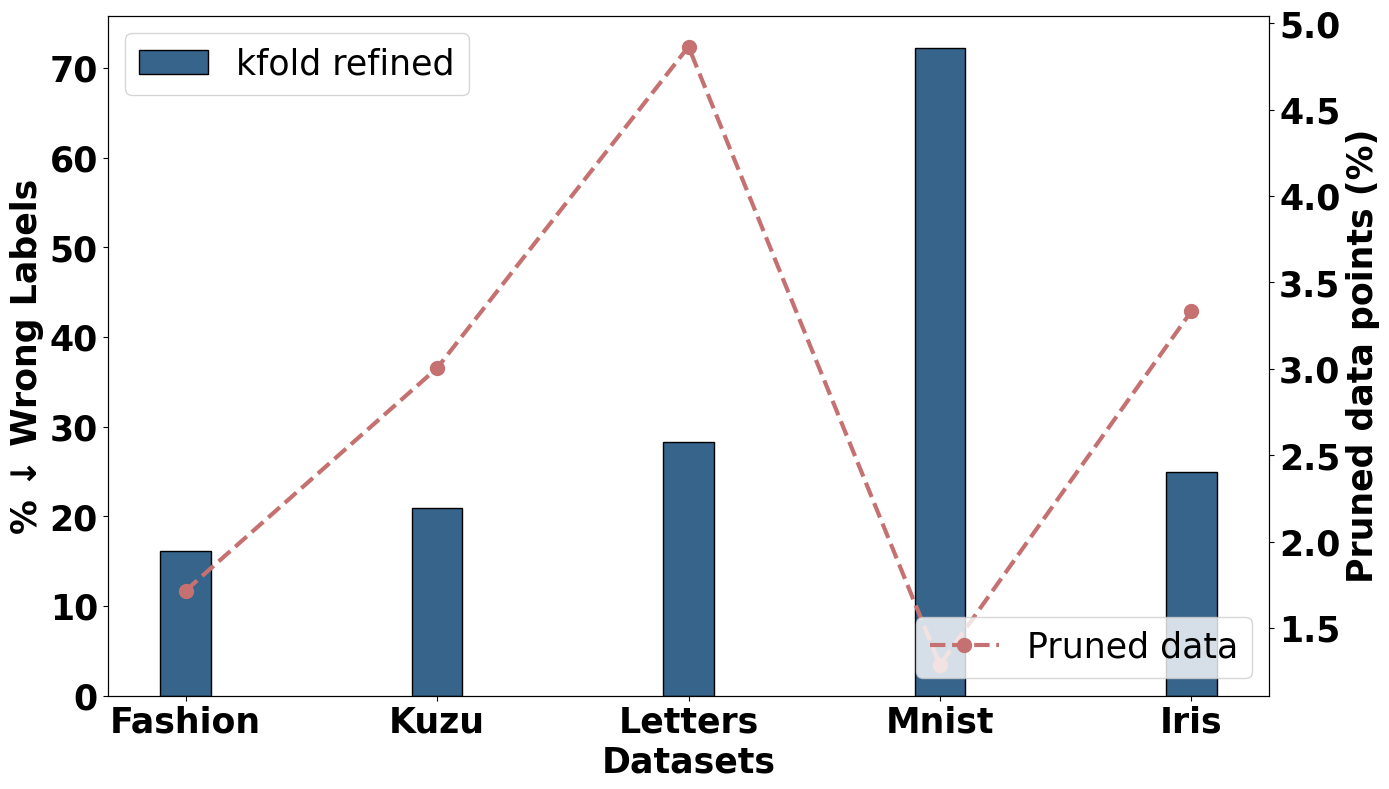}
    \caption{Comparison of k-fold dataset refining method (QNN models trained on the noisy Fake27QPulseV1 backend for 20 epochs with the final accuracy of $\approx$ 90\%). We note reduction in wrong labels and the percentage of pruned data points.}
    \label{kfold}
\vspace{-4mm}
\end{figure}
\subsection{Modified cost function for label obfuscation}

We utilize a modified cost function that obscures the true labels by incorporating masking labels. This is achieved by combining the standard cross-entropy loss with an additional loss term that promotes incorrect label prediction if the measurement is taken from all qubits or any combination other than the user-selected qubits. The modified loss function is represented as:

\[
\mathcal{L}_{\text{total}} = \mathcal{L}_{\text{correct}} + \alpha \mathcal{L}_{\text{adversary}}
\]

where $\mathcal{L}_{\text{correct}}$ is the cross-entropy loss between the predicted labels and the true labels, $\mathcal{L}_{\text{adversary}}$ is the cross-entropy loss between the predicted labels and the adversarial labels, and $\alpha$ is a weighting factor.

\section{Results and evaluation}

\subsection{Setup}

\subsubsection{Device:} In this work, all noiseless experiments were conducted using the ``lightning.qubit" device on Pennylane. Due to the long queue times and limited availability of real quantum devices, we used the ``qiskit.aer" device for our noisy training. We used Qiskit's Fake Provider module, specifically the Fake27QPulseV1, a 27-qubit noisy simulator that mimics IBM's Hanoi system and includes real hardware-calibrated data.

\subsubsection{Dataset:} Considering the exponential simulation runtime increase with growing qubits and large images, we utilized a reduced feature set for our experiments \cite{wang2022qoc}. We focused on MNIST, Fashion, Kuzushiji, and Letters datasets, reducing their dimensions to 8 using a convolutional autoencoder \cite{alam2021quantum} (original dimensions were 28×28). For each dataset, we created a subset with 4 classes: MNIST-4 (classes 0, 1, 2, 3), Fashion-4 (classes 6, 7, 8, 9), Kuzushiji(Kuzu)-4 (classes 3, 5, 6, 9), and Letters-4 (classes 1, 2, 3, 4). Each subset included 1000 samples, with 700 used for training and 300 for testing. Additionally, we also use the Iris dataset, which has 4 features and 3 classes, consisting of 156 samples (90 for training and 66 for testing). 

\subsubsection{Training:} We utilized an 8-qubit QNN for the MNIST, Fashion, Kuzushiji, and Letters datasets, and a 4-qubit QNN for the Iris dataset. Each QNN includes a classical data loading circuit, a Parametric Quantum Circuit (PQC), and measurement operations. Classical features were embedded using angle encoding with the R\textit{Z} gate. For the PQC, we selected Strongly Entangling Layers (SEL) \cite{schuld2020circuit} to create strong entanglement and enhance trainability \cite{cai2015entanglement}. The number of layers was adjusted based on dataset classes and features: 6 SEL layers for the Iris dataset and 12 SEL layers for the others. All experiments were conducted with 1000 shots/trials. Measurement operations consisted of expectation value measurements in the Pauli-Z basis. Training hyperparameters were: 30 epochs, learning rate of $10^{-3}$, batch size of 32 (16 for Iris), and Adam optimizer. Training was performed on an Intel Core-i7-12700H CPU with 40GB of RAM.


\subsection{Simulation Results}

\subsubsection{Label determination accuracy:}
To evaluate the effectiveness of the proposed label determination heuristics, we conducted experiments across five datasets: Fashion, Kuzu, Letters, Mnist, and Iris, using QNN models trained on a noiseless simulator for 20 epochs, with final training accuracy depicted in Fig.\ref{correct_labels}. We observe that an adversary can accurately predict the labels if the training accuracy of the victim's model is sufficiently high (around or greater than 90\%, which is usually the case). Each heuristic performs well in determining the correct label, achieving average correct prediction accuracies of 86\% for Majority Voting, 90\% for Weighted Linear Voting, and 92\% for Weighted Exponential Voting (Fig.\ref{correct_labels}). \textit{However, the accuracy of adversarial label prediction is inherently limited by the training accuracy of the victim's model.}

\begin{figure}
    \centering
    \includegraphics[width= 3.45in]{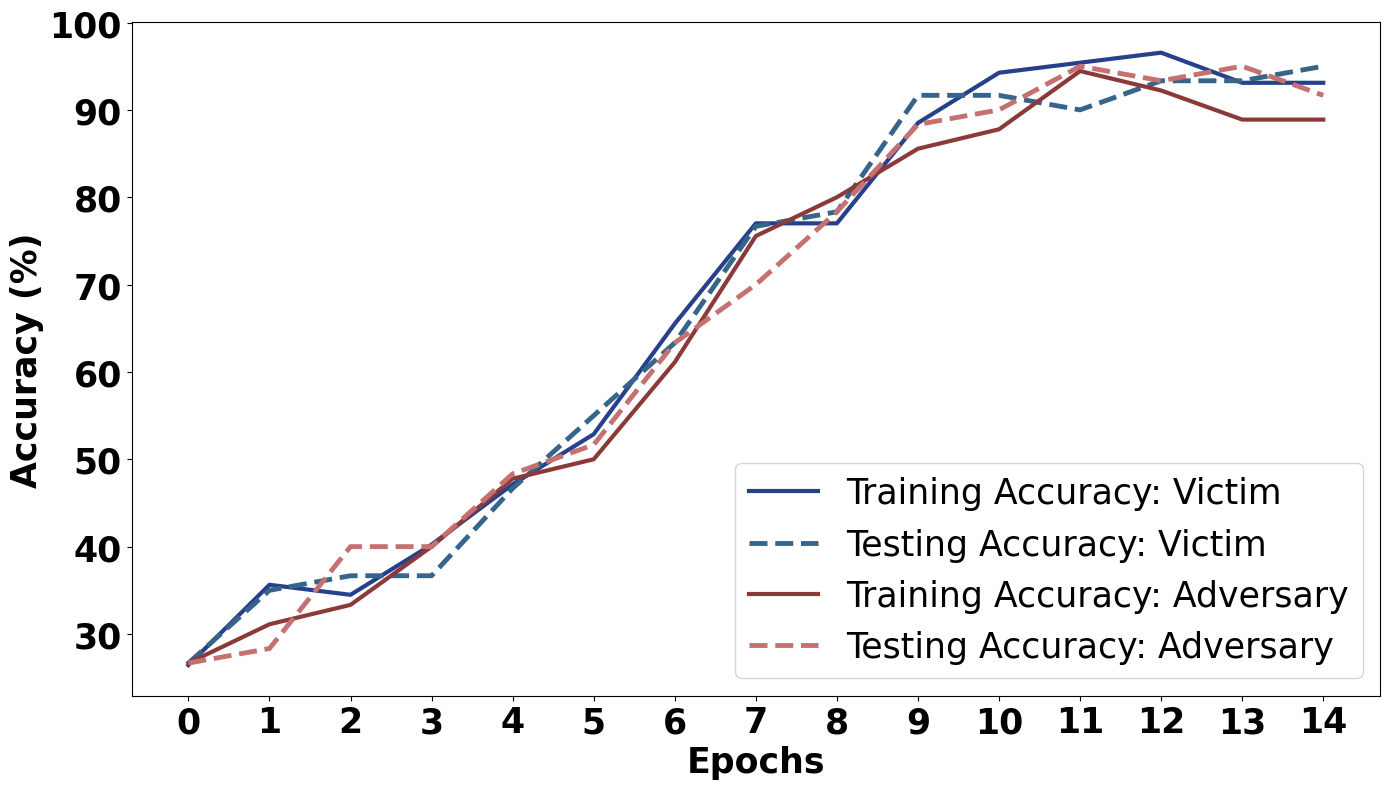}
    \caption{Training and testing accuracies for both the victim's original Iris dataset and the adversarially stolen and refined dataset using the Fake27QPulseV1 backend.}
    \label{attack}
\vspace{-4mm}
\end{figure}

\subsubsection{Training accuracy and label prediction:}

We investigated the correlation between adversarial label prediction accuracy and the final training accuracy of the victim's model using the Iris dataset. Experiments were conducted on the Fake27QPulseV1 backend device, training a QNN under noise with varying final training accuracies. We observed that all heuristics improve in label prediction as the training accuracy of the victim's QNN increases. As training accuracy reaches around 80\%, there is a noticeable improvement in prediction accuracy across all heuristics, with Weighted Exponential Voting consistently outperforming (Fig.\ref{iris_epochs}). On average, majority voting achieves 70\% accuracy, weighted linear voting achieves 76\%, and weighted exponential voting achieves 80\%. \textit{Weighted exponential voting performs better as it aligns with the training curve of the QML model.}

\begin{figure}
    \centering
    \includegraphics[width= 3.5in]{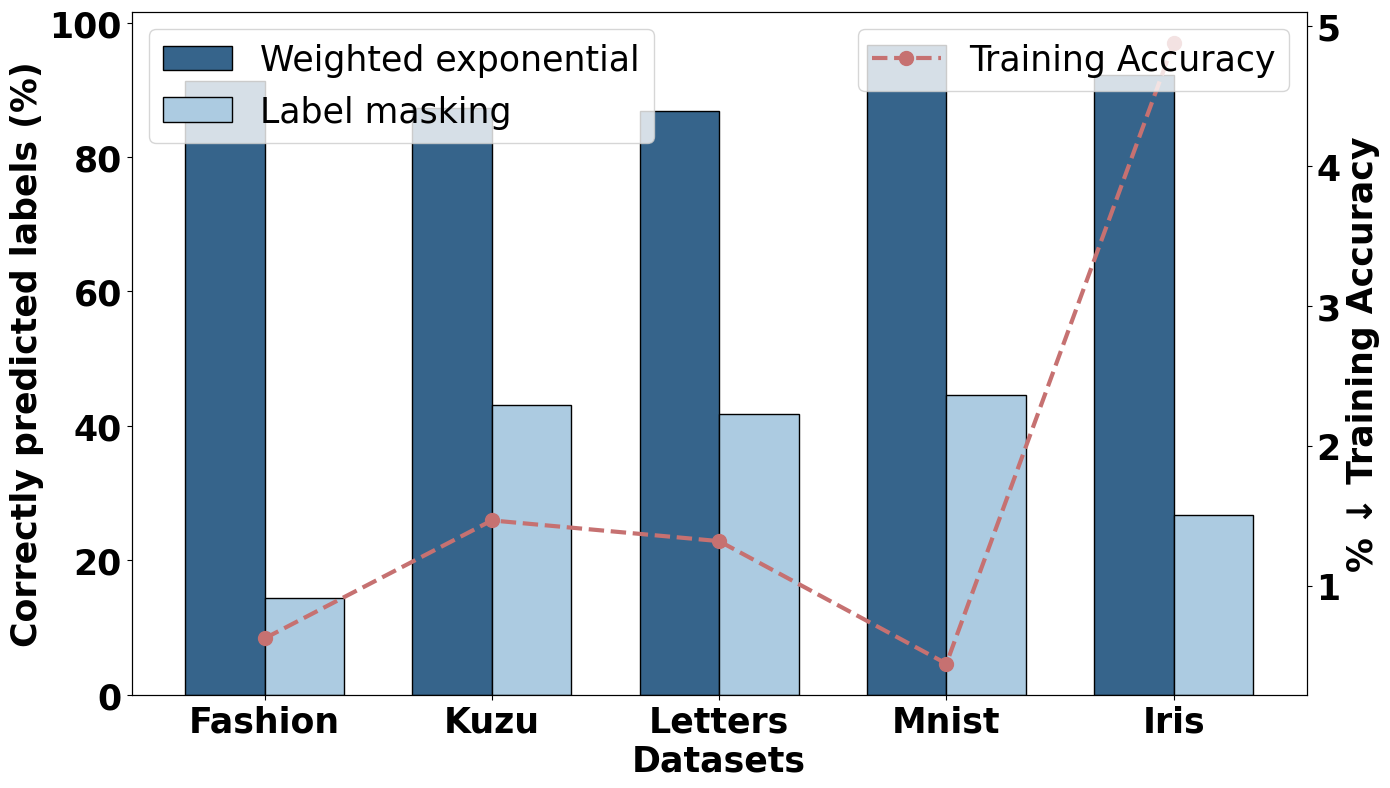}
    \caption{Correctly predicted label accuracy and training overhead for different datasets using proposed label masking methods. }
    \label{defense}
\vspace{-4mm}
\end{figure}

\subsubsection{Refining the dataset}

We conducted experiments across the five datasets to test the k-fold dataset refining method using QNN models trained on the noisy backend device Fake27QPulseV1 for 20 epochs (final training accuracy of $\approx$ 90\%). Baseline datasets were constructed using the superior weighted exponential voting method, and compared with iterative k-fold cross-validation method proposed to identify and correct/prune mislabeled data points. The plot in Fig. \ref{kfold} shows significant reductions in wrong labels: Fashion $\sim$18\%, Kuzu $\sim$22\%, Letters $\sim$35\%, Mnist $\sim$70\%, and Iris $\sim$20\%. The right y-axis indicates the percentage of pruned data points, with Fashion at $\sim$1.5\%, Kuzu at $\sim$2\%, Letters at $\sim$3.5\%, Mnist at $\sim$1.5\%, and Iris at $\sim$2\%. \textit{On average, the method achieved a 30\% reduction in wrong labels with only a $\sim$2\% average pruning of the data, highlighting that even with minimal pruning, a significant number of incorrect labels can be effectively removed.}

\subsubsection{Evaluating adversarial dataset performance}
We trained a 4-qubit QNN model using both the victim's original Iris dataset and the stolen and refined dataset by adversaries on the noisy backend Fake27QPulseV1. Both datasets exhibit similar performance over 15 epochs (Fig. \ref{attack}). The victim's (adversary's) dataset reaches a final training accuracy of approximately 93\% (90\%) and a testing accuracy of 88\% (86\%). Both datasets demonstrate consistent accuracy improvements across epochs, indicating effective learning. The minor differences in final accuracies can be attributed solely to the noise in the quantum circuits. \textit{These results suggest that the adversarial stolen and refined dataset can nearly replicate the performance of the victim's dataset.}

\subsubsection{Safeguarding training data}

We trained QNN models on five datasets: Fashion, Kuzu, Letters, Mnist, and Iris. For each dataset, we introduced a masking label, resulting in 4 classes (originally 3) for Iris and 5 (originally 4) for the others. The loss function parameter $\alpha$ was set to 1. Simulations for Iris were conducted on the noisy backend Fake27QPulseV1, while the others used a noiseless simulator. As shown in Fig. \ref{defense}, the weighted exponential method consistently achieves high label prediction accuracy across all datasets, nearing 100\% for Fashion, Kuzu, Letters, and Iris, and around 90\% for Mnist. Using the proposed defense methods: (1) introducing masking labels/classes and (2) a modified cost function—significantly reduces the adversary's label prediction accuracy. The performance varies across datasets: approximately 20\% for Fashion, 40\% for Kuzu, 30\% for Letters, 10\% for Mnist, and 25\% for Iris. On average, these methods reduce prediction accuracy by about 70\%. The proposed defense slightly impacts the user's training accuracy, with an average decrease of around 2\% over the same number of epochs. \textit{These results suggest that the label masking method effectively reduces the adversary's ability to predict correct labels, with only a minor impact on the user's training accuracy. By adjusting $\alpha$, users can further reduce the adversary's label prediction accuracy, though this will result in longer training times.}

\subsection{Discussion}

\subsubsection{Limitations of the Attack Model}

The success of the proposed attack model is limited by the training accuracy of the victim's QNN. If the model is poorly trained, the extracted data will be less reliable and valuable. Even with a successful attack, the adversary only obtains a dataset with some labels. They can identify distinct classes and correctly group data points, but cannot determine the actual nature of these classes. For instance, if the stolen dataset includes four classes, the adversary can identify the classification boundaries but without additional information cannot ascertain if these classes represent letters, animals, or other categories. Furthermore, our attack model is confined to scenarios using the most widely adopted last classical layer (softmax), used for converting expectation values into labels.


\subsubsection{Feature encoding}

In the problem of data encoding within PQCs, a critical challenge is whether an adversary can extract features or identify the user’s encoding method. In QML, techniques such as Principal Component Analysis (PCA), Independent Component Analysis (ICA), Linear Discriminant Analysis (LDA), Autoencoders, and Kernel methods are used to extract features from classical data due to qubit limitations, followed by embedding these features into quantum states. Typically, apart from the choice of encoding (e.g., angle or amplitude), features can be mapped to physical qubits in a round-robin fashion or other schemes, influencing the resultant quantum states. The proposed work focuses on the potential extraction of encoding circuits or, in simpler cases, angle-encoded features and associating labels with them. 
We assume that the adversary, 
knows the classical feature extraction method, type of embedding, and logical-to-physical mapping of data for the stolen state preparation circuit. This knowledge would enable them to host the cloned QML model and perform inferences correctly to make a profit. Determining the specific methods for such extraction remains an area for future research.

\section{Conclusion}
This work demonstrates exploitation of QNN's white-box access to adversaries in untrusted quantum cloud for extraction of encoded data and labels which can aid in subsequent model cloning attack. Simulations show that approximately 90\% of labels can be accurately extracted, and the cloned model achieves nearly the same accuracy as the original. To mitigate this threat, we propose masking labels/classes and modifying the cost function for label obfuscation. On average, these methods reduce adversarial prediction accuracy by about 70\%.

\section{Acknowledgment}

This work is supported in parts by NSF ( CNS-2129675, CCF-2210963, CCF-1718474, and DGE-2113839) and Intel's gift.

\bibliographystyle{ACM-Reference-Format}
\bibliography{sample-base}


\begin{thebibliography}{35}


\ifx \showCODEN    \undefined \def \showCODEN     #1{\unskip}     \fi
\ifx \showDOI      \undefined \def \showDOI       #1{#1}\fi
\ifx \showISBNx    \undefined \def \showISBNx     #1{\unskip}     \fi
\ifx \showISBNxiii \undefined \def \showISBNxiii  #1{\unskip}     \fi
\ifx \showISSN     \undefined \def \showISSN      #1{\unskip}     \fi
\ifx \showLCCN     \undefined \def \showLCCN      #1{\unskip}     \fi
\ifx \shownote     \undefined \def \shownote      #1{#1}          \fi
\ifx \showarticletitle \undefined \def \showarticletitle #1{#1}   \fi
\ifx \showURL      \undefined \def \showURL       {\relax}        \fi
\providecommand\bibfield[2]{#2}
\providecommand\bibinfo[2]{#2}
\providecommand\natexlab[1]{#1}
\providecommand\showeprint[2][]{arXiv:#2}

\bibitem[Abbas et~al\mbox{.}(2021)]%
        {abbas2021power}
\bibfield{author}{\bibinfo{person}{Amira Abbas}, \bibinfo{person}{David Sutter}, \bibinfo{person}{Christa Zoufal}, \bibinfo{person}{Aur{\'e}lien Lucchi}, \bibinfo{person}{Alessio Figalli}, {and} \bibinfo{person}{Stefan Woerner}.} \bibinfo{year}{2021}\natexlab{}.
\newblock \showarticletitle{The power of quantum neural networks}.
\newblock \bibinfo{journal}{\emph{Nature Computational Science}} \bibinfo{volume}{1}, \bibinfo{number}{6} (\bibinfo{year}{2021}), \bibinfo{pages}{403--409}.
\newblock


\bibitem[Alam et~al\mbox{.}(2021)]%
        {alam2021quantum}
\bibfield{author}{\bibinfo{person}{Mahabubul Alam}, \bibinfo{person}{Satwik Kundu}, \bibinfo{person}{Rasit~Onur Topaloglu}, {and} \bibinfo{person}{Swaroop Ghosh}.} \bibinfo{year}{2021}\natexlab{}.
\newblock \showarticletitle{Quantum-classical hybrid machine learning for image classification (iccad special session paper)}. In \bibinfo{booktitle}{\emph{2021 IEEE/ACM International Conference On Computer Aided Design (ICCAD)}}. IEEE, \bibinfo{pages}{1--7}.
\newblock


\bibitem[{Amazon}(2023)]%
        {MLaaS}
\bibfield{author}{\bibinfo{person}{{Amazon}}.} \bibinfo{year}{{2023}}\natexlab{}.
\newblock \bibinfo{title}{{Saiwa}}.
\newblock
\newblock
\urldef\tempurl%
\url{Saiwa. [a. n. d.]. Machine Learning as a Service (MLaaS) | Everything you need to know about that. https://saiwa.ai/blog/mlaas-1/}
\showURL{%
\tempurl}


\bibitem[{Amazon}(2024)]%
        {AmazonBraket}
\bibfield{author}{\bibinfo{person}{{Amazon}}.} \bibinfo{year}{{2024}}\natexlab{}.
\newblock \bibinfo{title}{{Amazon Braket}}.
\newblock
\newblock
\urldef\tempurl%
\url{https://aws.amazon.com/braket/}
\showURL{%
\tempurl}


\bibitem[Anguita et~al\mbox{.}(2012)]%
        {anguita2012k}
\bibfield{author}{\bibinfo{person}{Davide Anguita}, \bibinfo{person}{Luca Ghelardoni}, \bibinfo{person}{Alessandro Ghio}, \bibinfo{person}{Luca Oneto}, \bibinfo{person}{Sandro Ridella}, {et~al\mbox{.}}} \bibinfo{year}{2012}\natexlab{}.
\newblock \showarticletitle{The'K'in K-fold Cross Validation.}. In \bibinfo{booktitle}{\emph{ESANN}}, Vol.~\bibinfo{volume}{102}. \bibinfo{pages}{441--446}.
\newblock


\bibitem[Ash-Saki et~al\mbox{.}(2020)]%
        {ash2020analysis}
\bibfield{author}{\bibinfo{person}{Abdullah Ash-Saki}, \bibinfo{person}{Mahabubul Alam}, {and} \bibinfo{person}{Swaroop Ghosh}.} \bibinfo{year}{2020}\natexlab{}.
\newblock \showarticletitle{Analysis of crosstalk in nisq devices and security implications in multi-programming regime}. In \bibinfo{booktitle}{\emph{Proceedings of the ACM/IEEE International Symposium on Low Power Electronics and Design}}. \bibinfo{pages}{25--30}.
\newblock


\bibitem[{baidu}(2024)]%
        {baidu}
\bibfield{author}{\bibinfo{person}{{baidu}}.} \bibinfo{year}{{2024}}\natexlab{}.
\newblock \bibinfo{title}{{Quantum}}.
\newblock
\newblock
\urldef\tempurl%
\url{https://www.insidequantumtechnology.com/news-archive/chinas-baidu-rolls-beijing-based-quantum- computer-and-access-platform/}
\showURL{%
\tempurl}


\bibitem[Bauer et~al\mbox{.}(2020)]%
        {bauer2020quantum}
\bibfield{author}{\bibinfo{person}{Bela Bauer}, \bibinfo{person}{Sergey Bravyi}, \bibinfo{person}{Mario Motta}, {and} \bibinfo{person}{Garnet Kin-Lic Chan}.} \bibinfo{year}{2020}\natexlab{}.
\newblock \showarticletitle{Quantum algorithms for quantum chemistry and quantum materials science}.
\newblock \bibinfo{journal}{\emph{Chemical Reviews}} \bibinfo{volume}{120}, \bibinfo{number}{22} (\bibinfo{year}{2020}), \bibinfo{pages}{12685--12717}.
\newblock


\bibitem[Biamonte et~al\mbox{.}(2017)]%
        {biamonte2017quantum}
\bibfield{author}{\bibinfo{person}{Jacob Biamonte}, \bibinfo{person}{Peter Wittek}, \bibinfo{person}{Nicola Pancotti}, \bibinfo{person}{Patrick Rebentrost}, \bibinfo{person}{Nathan Wiebe}, {and} \bibinfo{person}{Seth Lloyd}.} \bibinfo{year}{2017}\natexlab{}.
\newblock \showarticletitle{Quantum machine learning}.
\newblock \bibinfo{journal}{\emph{Nature}} \bibinfo{volume}{549}, \bibinfo{number}{7671} (\bibinfo{year}{2017}), \bibinfo{pages}{195--202}.
\newblock


\bibitem[Busch et~al\mbox{.}(2014)]%
        {busch2014colloquium}
\bibfield{author}{\bibinfo{person}{Paul Busch}, \bibinfo{person}{Pekka Lahti}, {and} \bibinfo{person}{Reinhard~F Werner}.} \bibinfo{year}{2014}\natexlab{}.
\newblock \showarticletitle{Colloquium: Quantum root-mean-square error and measurement uncertainty relations}.
\newblock \bibinfo{journal}{\emph{Reviews of Modern Physics}} \bibinfo{volume}{86}, \bibinfo{number}{4} (\bibinfo{year}{2014}), \bibinfo{pages}{1261--1281}.
\newblock


\bibitem[{C. Q. Computing}(2021)]%
        {pytket}
\bibfield{author}{\bibinfo{person}{{C. Q. Computing}}.} \bibinfo{year}{{May 2021}}\natexlab{}.
\newblock \bibinfo{title}{{pytket}}.
\newblock
\newblock
\urldef\tempurl%
\url{https://cqcl.github.io/pytket/build/html/index.html}
\showURL{%
\tempurl}


\bibitem[Cai et~al\mbox{.}(2015)]%
        {cai2015entanglement}
\bibfield{author}{\bibinfo{person}{X-D Cai}, \bibinfo{person}{Dian Wu}, \bibinfo{person}{Z-E Su}, \bibinfo{person}{M-C Chen}, \bibinfo{person}{X-L Wang}, \bibinfo{person}{Li Li}, \bibinfo{person}{N-L Liu}, \bibinfo{person}{C-Y Lu}, {and} \bibinfo{person}{J-W Pan}.} \bibinfo{year}{2015}\natexlab{}.
\newblock \showarticletitle{Entanglement-based machine learning on a quantum computer}.
\newblock \bibinfo{journal}{\emph{Physical review letters}} \bibinfo{volume}{114}, \bibinfo{number}{11} (\bibinfo{year}{2015}), \bibinfo{pages}{110504}.
\newblock


\bibitem[Cao et~al\mbox{.}(2018)]%
        {cao2018potential}
\bibfield{author}{\bibinfo{person}{Yudong Cao}, \bibinfo{person}{Jhonathan Romero}, {and} \bibinfo{person}{Al{\'a}n Aspuru-Guzik}.} \bibinfo{year}{2018}\natexlab{}.
\newblock \showarticletitle{Potential of quantum computing for drug discovery}.
\newblock \bibinfo{journal}{\emph{IBM Journal of Research and Development}} \bibinfo{volume}{62}, \bibinfo{number}{6} (\bibinfo{year}{2018}), \bibinfo{pages}{6--1}.
\newblock


\bibitem[Cong et~al\mbox{.}(2019)]%
        {cong2019quantum}
\bibfield{author}{\bibinfo{person}{Iris Cong}, \bibinfo{person}{Soonwon Choi}, {and} \bibinfo{person}{Mikhail~D Lukin}.} \bibinfo{year}{2019}\natexlab{}.
\newblock \showarticletitle{Quantum convolutional neural networks}.
\newblock \bibinfo{journal}{\emph{Nature Physics}} \bibinfo{volume}{15}, \bibinfo{number}{12} (\bibinfo{year}{2019}), \bibinfo{pages}{1273--1278}.
\newblock


\bibitem[Farhi et~al\mbox{.}(2014)]%
        {farhi2014quantum}
\bibfield{author}{\bibinfo{person}{Edward Farhi}, \bibinfo{person}{Jeffrey Goldstone}, {and} \bibinfo{person}{Sam Gutmann}.} \bibinfo{year}{2014}\natexlab{}.
\newblock \showarticletitle{A quantum approximate optimization algorithm}.
\newblock \bibinfo{journal}{\emph{arXiv preprint arXiv:1411.4028}} (\bibinfo{year}{2014}).
\newblock


\bibitem[Fisher et~al\mbox{.}(2014)]%
        {fisher2014quantum}
\bibfield{author}{\bibinfo{person}{Kent~AG Fisher}, \bibinfo{person}{Anne Broadbent}, \bibinfo{person}{LK Shalm}, \bibinfo{person}{Z Yan}, \bibinfo{person}{Jonathan Lavoie}, \bibinfo{person}{Robert Prevedel}, \bibinfo{person}{Thomas Jennewein}, {and} \bibinfo{person}{Kevin~J Resch}.} \bibinfo{year}{2014}\natexlab{}.
\newblock \showarticletitle{Quantum computing on encrypted data}.
\newblock \bibinfo{journal}{\emph{Nature communications}} \bibinfo{volume}{5}, \bibinfo{number}{1} (\bibinfo{year}{2014}), \bibinfo{pages}{3074}.
\newblock


\bibitem[Ghosh et~al\mbox{.}(2023)]%
        {ghosh2023primer}
\bibfield{author}{\bibinfo{person}{Swaroop Ghosh}, \bibinfo{person}{Suryansh Upadhyay}, {and} \bibinfo{person}{Abdullah~Ash Saki}.} \bibinfo{year}{2023}\natexlab{}.
\newblock \showarticletitle{A primer on security of quantum computing}.
\newblock \bibinfo{journal}{\emph{arXiv preprint arXiv:2305.02505}} (\bibinfo{year}{2023}).
\newblock


\bibitem[Gilad-Bachrach et~al\mbox{.}(2016)]%
        {gilad2016cryptonets}
\bibfield{author}{\bibinfo{person}{Ran Gilad-Bachrach}, \bibinfo{person}{Nathan Dowlin}, \bibinfo{person}{Kim Laine}, \bibinfo{person}{Kristin Lauter}, \bibinfo{person}{Michael Naehrig}, {and} \bibinfo{person}{John Wernsing}.} \bibinfo{year}{2016}\natexlab{}.
\newblock \showarticletitle{Cryptonets: Applying neural networks to encrypted data with high throughput and accuracy}. In \bibinfo{booktitle}{\emph{International conference on machine learning}}. PMLR, \bibinfo{pages}{201--210}.
\newblock


\bibitem[{Google Quantum AI}(2024)]%
        {GoogleQuantumComputer}
\bibfield{author}{\bibinfo{person}{{Google Quantum AI}}.} \bibinfo{year}{{2024}}\natexlab{}.
\newblock \bibinfo{title}{{Google Quantum Computer}}.
\newblock
\newblock
\urldef\tempurl%
\url{https://quantumai.google/quantumcomputer}
\showURL{%
\tempurl}


\bibitem[Gupta et~al\mbox{.}(2023)]%
        {gupta2023quantum}
\bibfield{author}{\bibinfo{person}{Shivam Gupta}, \bibinfo{person}{Sachin Modgil}, \bibinfo{person}{Priyanka~C Bhatt}, \bibinfo{person}{Charbel Jose~Chiappetta Jabbour}, {and} \bibinfo{person}{Sachin Kamble}.} \bibinfo{year}{2023}\natexlab{}.
\newblock \showarticletitle{Quantum computing led innovation for achieving a more sustainable Covid-19 healthcare industry}.
\newblock \bibinfo{journal}{\emph{Technovation}}  \bibinfo{volume}{120} (\bibinfo{year}{2023}), \bibinfo{pages}{102544}.
\newblock


\bibitem[{IBM}(2024)]%
        {IBMQuantum}
\bibfield{author}{\bibinfo{person}{{IBM}}.} \bibinfo{year}{{2024}}\natexlab{}.
\newblock \bibinfo{title}{{IBM Quantum}}.
\newblock
\newblock
\urldef\tempurl%
\url{https://www.ibm.com/quantum}
\showURL{%
\tempurl}


\bibitem[Iverson and Preskill(2020)]%
        {iverson2020coherence}
\bibfield{author}{\bibinfo{person}{Joseph~K Iverson} {and} \bibinfo{person}{John Preskill}.} \bibinfo{year}{2020}\natexlab{}.
\newblock \showarticletitle{Coherence in logical quantum channels}.
\newblock \bibinfo{journal}{\emph{New Journal of Physics}} \bibinfo{volume}{22}, \bibinfo{number}{7} (\bibinfo{year}{2020}), \bibinfo{pages}{073066}.
\newblock


\bibitem[Kundu and Ghosh(2024)]%
        {kundu2024stiq}
\bibfield{author}{\bibinfo{person}{Satwik Kundu} {and} \bibinfo{person}{Swaroop Ghosh}.} \bibinfo{year}{2024}\natexlab{}.
\newblock \showarticletitle{STIQ: Safeguarding Training and Inferencing of Quantum Neural Networks from Untrusted Cloud}.
\newblock \bibinfo{journal}{\emph{arXiv preprint arXiv:2405.18746}} (\bibinfo{year}{2024}).
\newblock


\bibitem[Lloyd and Weedbrook(2018)]%
        {lloyd2018quantum}
\bibfield{author}{\bibinfo{person}{Seth Lloyd} {and} \bibinfo{person}{Christian Weedbrook}.} \bibinfo{year}{2018}\natexlab{}.
\newblock \showarticletitle{Quantum generative adversarial learning}.
\newblock \bibinfo{journal}{\emph{Physical review letters}} \bibinfo{volume}{121}, \bibinfo{number}{4} (\bibinfo{year}{2018}), \bibinfo{pages}{040502}.
\newblock


\bibitem[Magesan et~al\mbox{.}(2012)]%
        {magesan2012efficient}
\bibfield{author}{\bibinfo{person}{Easwar Magesan}, \bibinfo{person}{Jay~M Gambetta}, \bibinfo{person}{Blake~R Johnson}, \bibinfo{person}{Colm~A Ryan}, \bibinfo{person}{Jerry~M Chow}, \bibinfo{person}{Seth~T Merkel}, \bibinfo{person}{Marcus~P Da~Silva}, \bibinfo{person}{George~A Keefe}, \bibinfo{person}{Mary~B Rothwell}, \bibinfo{person}{Thomas~A Ohki}, {et~al\mbox{.}}} \bibinfo{year}{2012}\natexlab{}.
\newblock \showarticletitle{Efficient measurement of quantum gate error by interleaved randomized benchmarking}.
\newblock \bibinfo{journal}{\emph{Physical review letters}} \bibinfo{volume}{109}, \bibinfo{number}{8} (\bibinfo{year}{2012}), \bibinfo{pages}{080505}.
\newblock


\bibitem[Or{\'u}s et~al\mbox{.}(2019)]%
        {orus2019quantum}
\bibfield{author}{\bibinfo{person}{Rom{\'a}n Or{\'u}s}, \bibinfo{person}{Samuel Mugel}, {and} \bibinfo{person}{Enrique Lizaso}.} \bibinfo{year}{2019}\natexlab{}.
\newblock \showarticletitle{Quantum computing for finance: Overview and prospects}.
\newblock \bibinfo{journal}{\emph{Reviews in Physics}}  \bibinfo{volume}{4} (\bibinfo{year}{2019}), \bibinfo{pages}{100028}.
\newblock


\bibitem[Peng et~al\mbox{.}(2023)]%
        {peng2023rrnet}
\bibfield{author}{\bibinfo{person}{Hongwu Peng}, \bibinfo{person}{Shanglin Zhou}, \bibinfo{person}{Yukui Luo}, \bibinfo{person}{Nuo Xu}, \bibinfo{person}{Shijin Duan}, \bibinfo{person}{Ran Ran}, \bibinfo{person}{Jiahui Zhao}, \bibinfo{person}{Shaoyi Huang}, \bibinfo{person}{Xi Xie}, \bibinfo{person}{Chenghong Wang}, {et~al\mbox{.}}} \bibinfo{year}{2023}\natexlab{}.
\newblock \showarticletitle{Rrnet: Towards relu-reduced neural network for two-party computation based private inference}.
\newblock \bibinfo{journal}{\emph{arXiv preprint arXiv:2302.02292}} (\bibinfo{year}{2023}).
\newblock


\bibitem[Rebentrost et~al\mbox{.}(2014)]%
        {rebentrost2014quantum}
\bibfield{author}{\bibinfo{person}{Patrick Rebentrost}, \bibinfo{person}{Masoud Mohseni}, {and} \bibinfo{person}{Seth Lloyd}.} \bibinfo{year}{2014}\natexlab{}.
\newblock \showarticletitle{Quantum support vector machine for big data classification}.
\newblock \bibinfo{journal}{\emph{Physical review letters}} \bibinfo{volume}{113}, \bibinfo{number}{13} (\bibinfo{year}{2014}), \bibinfo{pages}{130503}.
\newblock


\bibitem[Schuld et~al\mbox{.}(2020)]%
        {schuld2020circuit}
\bibfield{author}{\bibinfo{person}{Maria Schuld}, \bibinfo{person}{Alex Bocharov}, \bibinfo{person}{Krysta~M Svore}, {and} \bibinfo{person}{Nathan Wiebe}.} \bibinfo{year}{2020}\natexlab{}.
\newblock \showarticletitle{Circuit-centric quantum classifiers}.
\newblock \bibinfo{journal}{\emph{Physical Review A}} \bibinfo{volume}{101}, \bibinfo{number}{3} (\bibinfo{year}{2020}), \bibinfo{pages}{032308}.
\newblock


\bibitem[Tilly et~al\mbox{.}(2022)]%
        {tilly2022variational}
\bibfield{author}{\bibinfo{person}{Jules Tilly}, \bibinfo{person}{Hongxiang Chen}, \bibinfo{person}{Shuxiang Cao}, \bibinfo{person}{Dario Picozzi}, \bibinfo{person}{Kanav Setia}, \bibinfo{person}{Ying Li}, \bibinfo{person}{Edward Grant}, \bibinfo{person}{Leonard Wossnig}, \bibinfo{person}{Ivan Rungger}, \bibinfo{person}{George~H Booth}, {et~al\mbox{.}}} \bibinfo{year}{2022}\natexlab{}.
\newblock \showarticletitle{The variational quantum eigensolver: a review of methods and best practices}.
\newblock \bibinfo{journal}{\emph{Physics Reports}}  \bibinfo{volume}{986} (\bibinfo{year}{2022}), \bibinfo{pages}{1--128}.
\newblock


\bibitem[Upadhyay and Ghosh(2022)]%
        {upadhyay2022robust}
\bibfield{author}{\bibinfo{person}{Suryansh Upadhyay} {and} \bibinfo{person}{Swaroop Ghosh}.} \bibinfo{year}{2022}\natexlab{}.
\newblock \showarticletitle{Robust and secure hybrid quantum-classical computation on untrusted cloud-based quantum hardware}. In \bibinfo{booktitle}{\emph{Proceedings of the 11th International Workshop on Hardware and Architectural Support for Security and Privacy}}. \bibinfo{pages}{45--52}.
\newblock


\bibitem[Upadhyay et~al\mbox{.}(2023)]%
        {upadhyay2023trustworthy}
\bibfield{author}{\bibinfo{person}{Suryansh Upadhyay}, \bibinfo{person}{Rasit~Onur Topaloglu}, {and} \bibinfo{person}{Swaroop Ghosh}.} \bibinfo{year}{2023}\natexlab{}.
\newblock \showarticletitle{Trustworthy computing using untrusted cloud-based quantum hardware}.
\newblock \bibinfo{journal}{\emph{arXiv preprint arXiv:2305.01826}} (\bibinfo{year}{2023}).
\newblock


\bibitem[Wang et~al\mbox{.}(2022)]%
        {wang2022qoc}
\bibfield{author}{\bibinfo{person}{Hanrui Wang}, \bibinfo{person}{Zirui Li}, \bibinfo{person}{Jiaqi Gu}, \bibinfo{person}{Yongshan Ding}, \bibinfo{person}{David~Z Pan}, {and} \bibinfo{person}{Song Han}.} \bibinfo{year}{2022}\natexlab{}.
\newblock \showarticletitle{Qoc: quantum on-chip training with parameter shift and gradient pruning}. In \bibinfo{booktitle}{\emph{Proceedings of the 59th ACM/IEEE Design Automation Conference}}. \bibinfo{pages}{655--660}.
\newblock


\bibitem[Wang et~al\mbox{.}(2023)]%
        {wang2023qumos}
\bibfield{author}{\bibinfo{person}{Zhepeng Wang}, \bibinfo{person}{Jinyang Li}, \bibinfo{person}{Zhirui Hu}, \bibinfo{person}{Blake Gage}, \bibinfo{person}{Elizabeth Iwasawa}, {and} \bibinfo{person}{Weiwen Jiang}.} \bibinfo{year}{2023}\natexlab{}.
\newblock \showarticletitle{Qumos: A framework for preserving security of quantum machine learning model}. In \bibinfo{booktitle}{\emph{2023 IEEE International Conference on Quantum Computing and Engineering (QCE)}}, Vol.~\bibinfo{volume}{1}. IEEE, \bibinfo{pages}{1089--1097}.
\newblock


\bibitem[{Z. Computing}(2021)]%
        {Orquestra}
\bibfield{author}{\bibinfo{person}{{Z. Computing}}.} \bibinfo{year}{{May 2021}}\natexlab{}.
\newblock \bibinfo{title}{{Orquestra}}.
\newblock
\newblock
\urldef\tempurl%
\url{https://www.zapatacomputing.com/orquestra/}
\showURL{%
\tempurl}


\end{thebibliography}
\end{document}